\documentclass[letterpaper,colorlinks=true,allcolors=ThemeColor,hyperfootnotes=false,11pt]{article}
\usepackage{ifluatex}

\newcommand{\affiliation}{Department of Economics, Tilburg University. Email: \href{mailto:y.takahashi@uvt.nl}{\texttt{y.takahashi@uvt.nl}}}

\usepackage[margin=1in]{geometry}

\ifluatex 
\usepackage{fontspec}
\usepackage{polyglossia}
\setdefaultlanguage{english} 
\else
\usepackage{lmodern} 
\usepackage[american]{babel}
\usepackage[utf8]{inputenc}
\usepackage[T1]{fontenc}
\fi 
\usepackage{newunicodechar}
\newunicodechar{€}{\texteuro}
\usepackage{microtype}
\usepackage{bookmark}
\usepackage[onehalfspacing]{setspace}
\usepackage[bottom]{footmisc}

\usepackage{graphicx}
\usepackage{rotating}
\usepackage[position=top,font=large]{subfig}
\usepackage[font=large]{caption}
\usepackage{pdflscape}
\usepackage{longtable}
\usepackage{threeparttable}
\usepackage{adjustbox}
\usepackage{multirow}
\usepackage{morefloats}
\usepackage{float}
\usepackage{boldline}
\usepackage{tabularx,booktabs}
\usepackage{rotating}
\usepackage{makecell}
\usepackage{caption}

\usepackage{pifont}
\newcommand{\cmark}{\ding{51}}%

\usepackage{comment}
\usepackage{soul}

\usepackage{amsmath}
\usepackage{amsfonts}
\usepackage{amsthm}
\usepackage{amssymb}
\usepackage{bbm}
\usepackage{siunitx}

\usepackage{tikz}
\usetikzlibrary{positioning}

\usepackage{csquotes}
\usepackage[backend=biber,authordate,uniquename=false,uniquelist=false,url=false,doi=false,isbn=false]{biblatex-chicago}
\AtEveryBibitem{
    \clearfield{urlyear}
    \clearfield{urlmonth}
    \clearfield{month}
    \clearfield{day}
    \clearfield{note}
    \clearlist{language}
}
\DeclareNosort{
  \nosort{type_names}{\regexp{[\x{2bf}\x{2018}]}}
}
\DeclareNoinit{
  \noinit{\regexp{[\x{2bf}\x{2018}]}}
}
\DeclareNameWrapperFormat{sortname}{\mkbibbold{#1}}
\DeclareNameWrapperAlias{author}{sortname}
\DeclareNameWrapperAlias{editor}{sortname}
\DeclareNameWrapperAlias{translator}{sortname}
\usepackage{xcolor}
\definecolor{ThemeColor}{RGB}{34,110,147} 

\usepackage[titletoc]{appendix}
\usepackage{pdfpages}
\usepackage[normalem]{ulem}
\usepackage{xurl}
\usepackage{enumitem}
\setlist{nosep}

\newtheorem*{pred*}{Prediction}

\makeatletter
\def\th@plain{%
  \thm@notefont{}
  \itshape 
}
\def\th@definition{%
  \thm@notefont{}
  \normalfont 
}
\makeatother

\usepackage{zref-xr,zref-user}
\usepackage{nameref}
\zxrsetup{toltxlabel}

\zexternaldocument*{CareerProgressionApp}
\urldef{\PAP}\url{https://osf.io/r6d8f/files}
\newcommand{\starnote}{Significance levels: * 10\%, ** 5\%, and *** 1\%}
\newcommand{\baselinenote}{Baseline mean and standard deviation are that of lower-IQ male receivers}
\usepackage{fancyhdr}
\fancypagestyle{PAP}{%
\fancyhf{}
\fancyhead[L]{Pre-analysis plan}
\fancyfoot[L]{\thepage}
}%
\fancypagestyle{Instructions}{%
\fancyhf{}
\fancyhead[L]{Experimental instructions: Main experiment}
\fancyfoot[L]{\thepage}
}%
\fancypagestyle{InstructionsFollowup}{%
\fancyhf{}
\fancyhead[L]{Experimental instructions: follow-up experiment}
\fancyfoot[L]{\thepage}
}%

\title{Are Men Less Generous to a Smarter Woman? Evidence from a Dictator Game Experiment}
\author{Yuki Takahashi\thanks{\affiliation.
I am grateful to Maria Bigoni, Boon Han Koh, Boris van Leeuwen, Natalia Montinari, and Bertil Tungodden for their invaluable feedback. I thank Editor Astrid Hopfensitz, two anonymous referees, and participants at the Applied Young Economist Webinar, the BEEN Meeting, and seminars at Ca' Foscari University, NHH, and the University of Bologna for their insightful comments and suggestions. Lorenzo Golinelli provided excellent technical and administrative assistance. The main experiment was conducted in compliance with the University of Bologna's standardized ethics protocol and personal data protection policy and was pre-registered with the OSF registry (\PAP). The Online Appendix A details the deviations from the pre-analysis plan. The follow-up experiment was approved by the Institutional Review Board of the Tilburg School of Economics and Management (approval no. IRB EXE 2023-044). Replication materials for this study are available at: \url{https://doi.org/10.17605/OSF.IO/YBE4M}.
}
}
\begin{document}
\maketitle
\begin{abstract} 
\noindent
Although evidence suggests men are more generous to women than to men, it may stem from paternalism and could reverse when women excel in important skills for one's career success, such as cognitive skills. Using a dictator game, this paper studies whether male dictators allocate less to female receivers than to male receivers when these receivers have higher IQs than dictators. By exogenously varying the receivers' IQ relative to the dictators', I do not find evidence consistent with this hypothesis; if anything, male dictators allocate slightly more to female receivers with higher IQs than to male receivers with equivalent IQs. The results hold both in mean and distribution and are robust to the so-called ``beauty premium.'' Also, female dictators' allocations are qualitatively similar to male dictators. These findings suggest that women who excel in cognitive skills may not receive less favorable treatment than equally intelligent men in the labor market. 
\end{abstract}

\newpage
\clearpage
\begin{flushright}
\textit{``All through my life, culturally reinforced signals cautioned me against being branded as too smart or too successful.''} \\
--- Sheryl Sandberg \textit{Lean In: Women, Work, and the Will to Lead}
\end{flushright}

\section{Introduction}
Cognitive skills are important personal attributes that affect one’s career success \parencite{herrnstein_bell_1996}. However, individual skills are not the only determinant of one's success: a positive and welcoming environment is also important, especially for jobs based on teamwork. Yet, colleagues' support may also be affected by gender norms, which makes the role of cognitive skills and its interplay with gender possibly more nuanced \parencite{eagly_role_2002,heilman_description_2001,ridgeway_gender_2001,rudman_backlash_2008}. Although the literature finds that men are more generous to women than to men \parencite{dufwenberg_generosity_2006,list_nature_2004}, few studies examine its interaction with cognitive skills.

One manifestation of gender norms is paternalism towards women, which may motivate men to be more generous to them. For example, many men (and women) in the US believe that women who fall behind deserve more support than men in similar situations because they attribute women's setbacks to bad luck rather than a lack of effort \parencite{cappelen_experimental_2023}. Similarly, men (and women) are more likely to attribute female leaders' bad outcomes to bad luck rather than poor decisions, compared to how they attribute male leaders' bad outcomes \parencite{erkal_women_2023}. Furthermore, many countries prohibit women from engaging in certain activities, such as close military combat \parencite{fitriani_women_2016}.\footnote{In fact, most countries exhibit protective attitudes toward women \parencite{glick_beyond_2000} and such attitudes are more pronounced in low-income countries. For example, \textcite{buchmann_paternalistic_2024} show that men prohibit women from taking risky jobs against their will.} 

The main research question addressed in this paper is whether men are less generous to women who have higher cognitive skills than they do. I hypothesize that this is indeed the case because if the paternalism derived from gender norms drives men's higher generosity towards women, their generosity may disappear or even reverse when women violate these norms by excelling in critical skills for their career success. I test this hypothesis via a laboratory experiment.

In the experiment, participants first work on an incentivized IQ test (Raven matrices) that measures their cognitive skills. After the test, participants are randomly assigned to a group of six and receive an IQ rank relative to other group members. Then three of the six members are randomly chosen to be dictators and play three rounds of dictator game with the other three members chosen to be receivers, observing the receivers' facial photos and first names -- both of which convey information about their gender -- and the relative IQ ranks. I use the dictators' allocation as the measure of generosity, a widely used tool in experimental economics and shown to predict one's generosity outside the laboratory \parencite{franzen_external_2013}. The use of photos allows me to inform the dictators of the receivers' gender with minimum experimenter demand effects.\footnote{For example, \textcite{babcock_gender_2017}, \textcite{coffman_evidence_2014}, and \textcite{isaksson_it_2018} use photos to inform experimental participants of the other participants' gender. Yet, I show the robustness of the results to the concern that facial photos may be subject to the so-called ``beauty premium.''} I use dictator IQ fixed effects in the analysis to compare allocations of dictators with the same IQ but assigned different relative IQ ranks due to random group formation to cut the correlation between IQ and baseline generosity.\footnote{IQ fixed effects were used by \textcite{zimmermann_dynamics_2020}, among others.}\textsuperscript{,}\footnote{For example, \textcite{falk_socioeconomic_2021} show that children from families with higher socioeconomic status tend to have higher IQ and are more altruistic than children from families with lower socioeconomic status.}

I first confirm that male and female dictators' behaviors are consistent with the literature: their allocation amounts are similar to those in a previous study with a comparable level of social distance from receivers. Male dictators allocate more to female receivers than to male receivers (albeit statistically insignificant), and female dictators allocate more than male dictators. However, I do not find evidence that male dictators are less generous to female receivers with higher IQs. The point estimate is quantitatively negligible, statistically indistinguishable from zero, and has a tight confidence interval. If anything, male dictators are slightly more generous to female receivers with higher IQs. These results are robust to the so-called ``beauty premium'' and hold across the entire distribution of male dictator allocations. The allocation patterns of female dictators are qualitatively similar to those of male dictators. Taken together, contrary to the hypothesis, men (and women) are no less generous to women, even when women excel in cognitive skills. Thus, the results suggest that women who excel in cognitive skills may not receive less favorable treatment than equally intelligent men in the labor market.\footnote{A caveat is that the size of the standard error is not very small; thus, the results should be interpreted with caution.}

This paper contributes to the literature about people's differential attitudes towards competent women and men. Social psychology and sociology literature find that people perceive and evaluate competent women more negatively than competent men \parencite{heilman_penalties_2004,phelan_competent_2008,rudman_self-promotion_1998,rudman_reactions_2004,rudman_status_2012}. Also, \textcite{quadlin_mark_2018} finds female college students with a very high grade point average (GPA) receive fewer callbacks in hiring than male students with a similar GPA. However, there are studies that find the opposite: \textcite{ceci_women_2015} and \textcite{williams_national_2015} find well-qualified female candidates for assistant professor positions receive equal or more favorable treatment than equally qualified male candidates in hiring. I show that although intelligence is an important factor of career success and one of the attributes men (and women) care most about \parencite{castagnetti_protecting_2022,eil_good_2011,zimmermann_dynamics_2020}, excelling in intelligence is not a sufficient condition for women to receive less favorable treatments than equally intelligent men, consistent with the latter line of studies.

My paper also contributes to the literature on the role of gender in dictator games. The literature finds that female dictators allocate more than male dictators, but the difference is quantitatively modest at best \parencite{bilen_are_2021}. The difference is also context-dependent \parencite{croson_gender_2009,donate-buendia_gender_2022} and possibly driven by non-monetary motives \parencite{klinowski_gender_2018}. Regarding the gender of the receivers, the literature finds that dictators allocate more to female receivers than to male receivers \parencite{engel_dictator_2011}, and male dictators may do so more \parencite{dufwenberg_generosity_2006,list_nature_2004}. My paper adds to the latter evidence by introducing IQ as an additional dimension to gender and shows that male dictators' allocation patterns are largely unchanged when female receivers have higher IQs than male dictators.

The remainder of the paper proceeds as follows. Section \ref{sec:Experiment} describes the experimental design, procedure, implementation, and data. Section \ref{sec:EmpStrategy} discusses the empirical strategy. Section \ref{sec:Results} presents the results. Section \ref{sec:Conclusion} concludes.

\section{Experiment}\label{sec:Experiment}
I conduct two experiments: the main and the follow-up. The main experiment collects data on dictator game allocation, and the follow-up experiment collects data on the receivers' attractiveness and other facial characteristics. I describe each of them in detail below.

\subsection{Main Experiment}
The main experiment consists of two parts, and participants receive instructions at the beginning of each part.

\subsubsection*{Pre-Experiment: Random Desk Assignment \& Photo-Taking}
Participants are randomly assigned to a partitioned computer desk. Afterward, they have their facial photos taken at a photo booth and enter their first names on their computers. The experimenters then go to each participant's desk to check that their photo and first name are correct, to assure other participants that the checked participants' photos and first names are real, following \textcite{isaksson_it_2018}.

\subsubsection*{Part 1: IQ Test}
In Part 1, participants work on nine incentivized IQ questions for nine minutes to measure their IQ. I use \textcite{bilker_development_2012}'s ``form A 9-item Raven test,'' which measures one's IQ 90\% as accurately as the full-length Raven test but with fewer questions. Participants receive 0.5\texteuro\ for each correctly solved IQ test question, but they do not receive information about how many questions they have solved correctly until the end of the experiment.

After the IQ test, participants make an incentivized guess on the number of IQ test questions they have solved correctly, which I use as a measure of their over-confidence level. They receive 0.5\texteuro\ for a correct guess. They do not receive feedback on their guess accuracy until the end of the experiment.

Following \textcite{eil_good_2011}, six participants are randomly grouped and privately informed about their IQ rank relative to the other five participants in the group. Ties are broken randomly. They then answer a set of comprehension questions about their IQ rank. They cannot proceed to the next part until they answer the comprehension questions correctly.

\subsubsection*{Part 2a: Dictator Game (Dictators Only)}

\begin{figure}[!htb]
\begin{center}
\caption{Dictator's allocation screen}\label{fig:DGPage}
\includegraphics[width=\columnwidth]{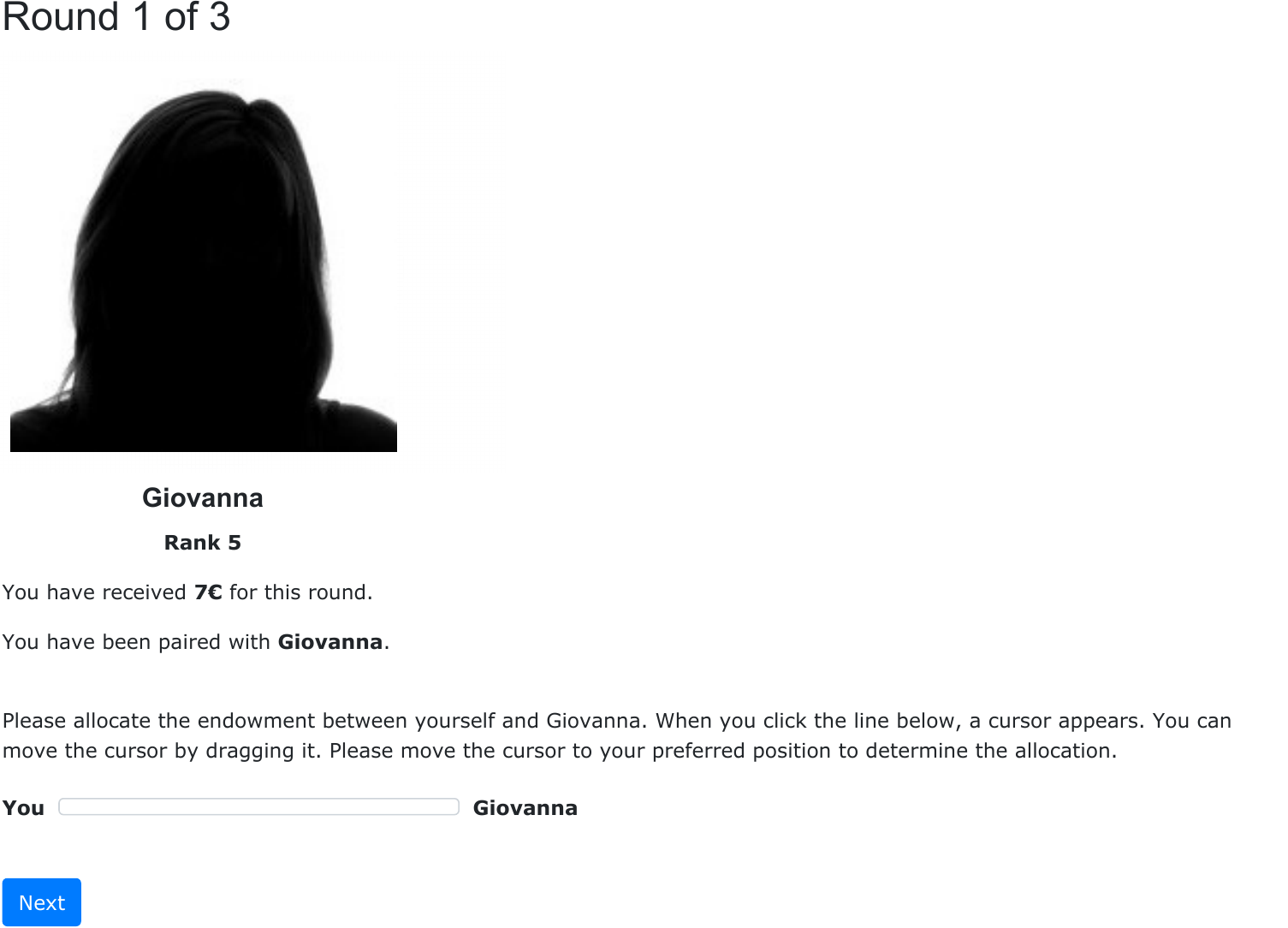}
\end{center}
\vspace{-12pt}
\begin{tablenotes}[flushleft]
\item \footnotesize \textit{Notes:} This figure shows an example of a dictator's allocation screen. In this example, the dictator is playing the first round and paired with a receiver whose first name is Giovanna (a female name) and whose IQ rank is 5. In the experiment, dictators see Giovanna's facial photo instead of the silhouette.
\end{tablenotes}
\end{figure}

In Part 2, three participants in each group are randomly assigned the role of dictators (which I call ``active participant''), and the other three participants are randomly assigned the role of receivers (which I call ``passive participant''). Dictators then play dictator games with windfall money three times, each time with a different receiver from their group, one by one, using a perfect stranger-matching protocol. During the dictator game, dictators observe the receivers' facial photos, first names, and IQ ranks; see Figure \ref{fig:DGPage} for an example of a dictator's decision screen. The use of photos follows gender economics literature \parencite{babcock_gender_2017,coffman_evidence_2014,isaksson_it_2018} to minimize the experimenter demand effects.

Dictators are told that their decisions are anonymous to the receivers and that their allocation will be paid as a ``top-up'' to the receivers' earnings. Dictators decide allocations by moving a cursor on a slider where the cursor is initially hidden to prevent anchoring. I vary the endowment across rounds to make each dictator game less repetitive: 7\texteuro\ for the 1st and the 3rd rounds, and 5\texteuro\ for the 2nd round. At the end of the experiment, one of the three allocations is randomly chosen for each participant as earnings for Part 2.\footnote{For each dictator and in each round, one of the three receivers in the same group is randomly chosen \textit{without replacement} and the dictator allocates the endowment between themselves and the receiver. Thus, it is possible for two dictators to play the dictator game with the same receiver in the same round. At the end of the dictator games, each participant has three allocations, one of which is randomly chosen for payment.}\textsuperscript{,}\footnote{Note that, as with other information provision experiments, the results I show later are intention-to-treat effects because some dictators may not have taken into account the receivers' IQ rank when deciding the allocation. Yet, dictators solved the comprehension questions about their IQ rank relative to other group members, as in Appendix Figure D1.}

\subsubsection*{Part 2b: Belief Elicitation (Receivers Only)}
I also collect a proxy of dictators' beliefs on how many IQ test questions the matched receivers have solved correctly. To prevent the belief elicitation from affecting/being affected by the dictator game, I exploit the random role assignment, and use the receivers' beliefs. This is a valid proxy because both dictators and receivers are exactly in the same experimental environment up to the role assignment and the role assignment is random. Specifically, while dictators are playing the dictator game, receivers are matched with the other two receivers in the same group one by one with a perfect stranger-matching protocol, and they make incentivized guesses on how many IQ test questions they have solved correctly, observing the receivers' facial photo, first name, and IQ rank, just as dictators do. They receive 0.5\texteuro\ for each correct guess.

To address the non-anonymity of showing facial photos and first names, I ask both dictators and receivers how well they know the paired participants on a scale of 4, from ``Did not know at all'' to ``Knew very well.''\footnote{See the experimental instructions in Appendix E for the exact wording.} I ask this question twice to make sure they do not answer randomly: right after the three dictator games (for dictators) or two guesses (for receivers) and in the post-experimental questionnaire.

\subsubsection*{Post-Experiment: Questionnaire}
After the dictator game and the belief elicitation, participants are told their earnings from the IQ test, the dictator game, and the belief elicitation in private. Before receiving their earnings, participants answer a short questionnaire about their demographics. In addition, receivers are asked whether I could use their photos in another experiment with a gratuity of 1.5\texteuro\ (only for receivers who agreed; 149 receivers out of 193, or 77\% of receivers, agreed).

\subsubsection*{Implementation}
The main experiment was programmed with oTree \parencite{chen_otree--open-source_2016} and conducted in English in November-December 2019 at the Bologna Laboratory for Experiments in Social Science (BLESS). I recruited 390 students (195 female and 195 male) of the University of Bologna via ORSEE \parencite{greiner_subject_2015} who (i) were born in Italy, (ii) had not participated in gender-related experiments in the past (as far as I could trace), and (iii) were available to participate in experiments in English. The first condition was to reduce the chance that receivers' photos and first names would signal ethnicity, race, or cultural background. The second condition was to minimize the experimenter demand effects. The third condition was to run the experiment in English. As a further attempt to make the data cleaner, I excluded receivers with non-Italian-sounding names and allocations in which the dictators answered at least once that they knew the paired receivers ``very well.''\footnote{Although it is easy to distinguish Italian and non-Italian-sounding names, to make sure not to misclassify I asked the laboratory manager, who was a native Italian, to check the participants' first names after each session.} These data screenings leave me with 388 participants, 195 dictators, and 558 dictator allocations.

The number of participants is based on the power simulation in the pre-analysis plan to achieve 80\% power.\footnote{I exclude the first-session data because of the problem discussed in Appendix A.} The experiment is pre-registered with the OSF.\footnote{The pre-analysis plan and the R code for power calculation are available at the OSF registry: \PAP.} Appendix A explains deviations from the pre-analysis plan.

I ran 24 sessions in total, and the number of participants in each session was a multiple of 6 (12 to 30). The average session length was 70 minutes, including registration and payment. The average payment per participant was about 10\texteuro\, including the participation fee of 2.5\texteuro\ and a gratuity of 1.5\texteuro\ for photo use in another experiment (only for receivers who agreed).

\subsection{Follow-Up Experiment}
I conduct a follow-up experiment to collect a measure of the attractiveness and other facial characteristics of the receivers in the main experiment. It consists of one part, and participants earn a flat fee of 10\texteuro.

Participants are randomly assigned to a partitioned computer desk. Afterward, participants see photos of 100 receivers in the main experiment one by one and rate the attractiveness of each of the photos on a 5-point Likert scale. I also ask them to rate the photos in terms of how kind they look (on a 5-point Likert scale), to which extent they look Italian (on a 3-point Likert scale), and whether they know the person (yes or no) -- all in one screen for each photo. The photos are randomly drawn from the pool of all receivers who agreed to show their photos in another experiment, and the order of the photos is randomized. After rating all the 100 photos, participants provide their gender and age.

The follow-up experiment was programmed with oTree and conducted in Italian in October 2023 at the Bologna Laboratory for Experiments in Social Science (BLESS). I recruited 28 students (14 female and 14 male) who (i) were born in Italy and (ii) were available to participate in experiments in English to make the subject pool as close as possible to that of the main experiment. I also restricted participants to those who had not participated in the main experiment.\footnote{Participants know the people in the photos in 10 out of 2800, or 0.4\% of total ratings.}\textsuperscript{,}\footnote{Although I recruited participants born in Italy, 5 people in the photos were considered completely non-Italian (4 by male raters and 4 by female raters with overlaps of 3 receivers).}\textsuperscript{,}\footnote{I use ratings by male raters for male dictators and by female raters for female dictators. For female and male raters, the median ratings per receiver is 9, the mean is 8.66, the minimum is 4, the maximum is 14, and the standard deviation is 1.84.}

\subsection{Data}\label{sec:DataDesc}
Appendix B describes the data in detail; I briefly summarize its key aspects here. First, most male dictators (95\%) do not know the receiver at all, so it is unlikely that the relationship outside the laboratory affects dictator allocation. Second, consistent with the literature, male dictators allocate less (6.3 percentage points less, the p-value < 0.05 with a two-sided t-test) to receivers than female dictators do. Third, again somewhat consistent with the literature, dictators, especially male dictators, allocate slightly more (2.4 percentage points more; male dictators allocate 3.1 percentage points more) to female receivers than to male receivers, although the difference is statistically insignificant. Fourth, both men and women consider women to look more attractive than men, regardless of their IQ level.

\section{Empirical Strategy}\label{sec:EmpStrategy}
A naive way to test whether male dictators allocate less to higher-IQ female receivers than to higher-IQ male receivers is to compare male dictators’ allocations to those receivers. However, this simple difference may be biased upward because male dictators may allocate more to more attractive receivers \parencite{rosenblat_beauty_2008}, and men rate women more highly than men on attractiveness (see the discussion in Section \ref{sec:DataDesc}). In fact, male dictators allocate more to more attractive receivers (see Appendix Figure D2, Panels A and B).\footnote{Interestingly, female dictators do not allocate more to more attractive receivers (see Appendix Figure D2, Panels C and D).}

A difference-in-differences would eliminate the bias from the attractiveness differences, using the difference between male dictators’ allocations to lower-IQ female receivers and to lower-IQ male receivers as a comparison group. However, it introduces two other biases because low-IQ dictators are more likely to face higher-IQ receivers. First, because low-IQ dictators earned less on the IQ test than high-IQ dictators, higher-IQ receivers and dictators’ earnings are negatively correlated, potentially biasing the dictators’ allocation to higher-IQ receivers due to the wealth effects. Also, if low-IQ dictators think the differences between their earnings and the higher-IQ receivers’ earnings are larger than high-IQ dictators think, it also induces inequality aversion \parencite{fehr_theory_1999}. Second, because one’s IQ is positively associated with one’s socioeconomic background, and one’s socioeconomic background influences one’s social preferences \parencite{falk_socioeconomic_2021}, the dictators’ allocation to higher-IQ receivers may be biased upward or downward. In my sample, male dictators’ IQs (or their earnings on the IQ test) and allocations are negatively correlated, and male dictators’ IQ ranks and allocations are positively correlated (see Appendix Figure D2, Panels A and B), both of which bias the allocations to higher-IQ receivers upwards.\footnote{We do not observe this correlation patterns for female dictators (see Appendix Figure D2, Panels C and D).}

To address these concerns, I estimate the following difference-in-differences equation via OLS using male dictators' allocation data:
\begin{equation}\label{eq:MainEq}
\begin{split}
Allocate_{ij} = &\beta_1 HigherIQReceiver_{ij}\times FemaleReceiver_j \\
&+ \beta_2 HigherIQReceiver_{ij} + \beta_3 FemaleReceiver_j + X_{ij}'\gamma + \mu_i^{IQ} + \epsilon_{ij}
\end{split}
\end{equation}
where each variable is defined as follows:
\begin{itemize}
\item $Allocate_{ij}\in [0,1]$: dictator $i$'s allocation to receiver $j$ as a fraction of endowment.
\item $HigherIQReceiver_{ij}\in \{0,1\}$: an indicator variable equals 1 if receiver $j$'s IQ is higher than that of dictator $i$. 
\item $FemaleReceiver_j \in \{0,1\}$: an indicator variable equals 1 if receiver $j$ is female.
\item $X_{ij}$: a set of additional covariates to increase statistical power. Appendix C provides a full description of the covariates.
\item $\epsilon_{ij}$: the error term.
\end{itemize}
and $\mu_i^{IQ} \equiv \sum_{k=1}^9 \theta_k^{IQ}\mathbbm{1}[\text{i's IQ}=k]$ is fixed effects for the dictators' IQ (the number of IQ test questions they have solved correctly), where $\mathbbm{1}$ is the indicator function. Standard errors are clustered at the dictator level with \textcite{pustejovsky_small-sample_2018}'s small cluster bias adjustment. The IQ fixed effects are included following \textcite{zimmermann_dynamics_2020}, so that the coefficients in equation \ref{eq:MainEq} capture allocation differences due to the receivers' IQ, not that of the dictators. As discussed above, without the IQ fixed effects, the coefficients related to higher-IQ receiver can capture the wealth effects, inequality aversion, and the difference in the preference for allocation between high IQ and low IQ dictators.

The coefficient of interest is $\beta_1$, which captures the difference between the dictator allocations to higher-IQ female receivers and higher-IQ male receivers relative to the difference between the dictator allocations to lower-IQ female receivers and lower-IQ male receivers. As discussed above, because men rate women more highly than men on attractiveness, the estimates related to the female receiver can capture the effects of attractiveness differences without taking the double differences. On the other hand, as discussed in Section \ref{sec:DataDesc}, men's ratings on high-IQ women's attractiveness and low-IQ women's attractiveness are very similar. Also, men's rating on high-IQ men's attractiveness and low-IQ men's attractiveness is very similar, justifying the difference-in-differences.

\section{Results}\label{sec:Results}
\subsection{Regression Results}\label{sec:RegResults}

\begin{table}[!htbp]
\begin{center}
\caption{Dictator allocations to higher-IQ female receivers -- OLS, male dictators}\label{tab:MainRegMale}
\begin{adjustbox}{max width=\textwidth}
\begin{tabular}{lcccc}
\toprule
Outcome: & \multicolumn{4}{c}{Dictator's allocation (fraction of endowment)} \\
\midrule
Sample: & \multicolumn{4}{c}{Male dictators}  \\
 & (1) & (2) & (3) & (4) \\
\midrule
Higher IQ receiver x Female receiver ($\beta_1$) & 0.018 & 0.017 & 0.031 & 0.047\\
 & (0.060) & (0.060) & (0.061) & (0.075)\\
 & {}[-0.101, 0.136] & {}[-0.101, 0.134] & {}[-0.089, 0.151] & {}[-0.101, 0.195]\\
Higher IQ receiver ($\beta_2$) & 0.093* & 0.054 & 0.048 & 0.007\\
 & (0.048) & (0.053) & (0.055) & (0.059)\\
 & {}[-0.001, 0.188] & {}[-0.050, 0.159] & {}[-0.060, 0.156] & {}[-0.109, 0.124]\\
Female receiver ($\beta_3$) & 0.038 & 0.031 & 0.014 & -0.021\\
 & (0.035) & (0.035) & (0.034) & (0.044)\\
 & {}[-0.031, 0.107] & {}[-0.038, 0.100] & {}[-0.053, 0.081] & {}[-0.108, 0.066]\\
\midrule
Dictator IQ FE             & - & \cmark & \cmark & \cmark \\
Round FE                   & - & -      & \cmark & \cmark \\
Social distance FE         & - & -      & \cmark & \cmark \\
Dictator demographics      & - & -      & \cmark & \cmark \\
Receiver demographics      & - & -      & \cmark & \cmark \\
Receiver attractiveness FE & - & -      & -      & \cmark \\
\midrule
\multirow{2}{*}{\makecell[l]{Higher IQ receiver x Female receiver\\\quad+Higher IQ receiver ($\beta_1+\beta_2$)}} & 0.111** & 0.071 & 0.079 & 0.054\\
 & (0.055) & (0.048) & (0.052) & (0.070)\\
 & {}[0.004, 0.219] & {}[-0.024, 0.166] & {}[-0.023, 0.181] & {}[-0.084, 0.193]\\
\multirow{2}{*}{\makecell[l]{Higher IQ receiver x Female receiver\\\quad+Female receiver ($\beta_1+\beta_3$)}} & 0.056 & 0.048 & 0.045 & 0.026\\
 & (0.049) & (0.046) & (0.047) & (0.061)\\
 & {}[-0.040, 0.151] & {}[-0.043, 0.139] & {}[-0.048, 0.138] & {}[-0.094, 0.146]\\
Baseline Mean & 0.305 & 0.305 & 0.305 & 0.327\\
Baseline SD & 0.269 & 0.269 & 0.269 & 0.270\\
Adj. R-squared & 0.032 & 0.052 & 0.080 & 0.088\\
Observations & 260 & 260 & 260 & 211\\
Clusters & 91 & 91 & 91 & 91\\
\bottomrule
\end{tabular}
\end{adjustbox}
\end{center}
\vspace{-12pt}
\begin{tablenotes}[flushleft]
\item \footnotesize \textit{Notes:} This table presents the regression results of equation \ref{eq:MainEq}. The standard error (in parenthesis) and the 95\% confidence interval (in bracket) are reported below each coefficient estimate. The standard errors are clustered at the dictator level with \textcite{pustejovsky_small-sample_2018}'s small cluster bias adjustment. \baselinenote. \starnote.
\end{tablenotes}
\end{table}%

Table \ref{tab:MainRegMale} presents the regression results of equation \ref{eq:MainEq} with data for male dictators. Column 1 shows that the coefficient estimate on higher-IQ receivers ($\beta_2$) is 0.093 and statistically marginally significant at 10\%, suggesting that male dictators allocate 9.3 percentage points more to higher-IQ male receivers than to lower-IQ male receivers. Also, the sum of the coefficient estimates on the interaction term between higher-IQ receivers and the female receivers, and on the higher-IQ receivers  ($\beta_1+\beta_2$) is 0.111 and statistically significant at 5\%, suggesting that male dictators allocate 11.1 percentage points more to higher-IQ female receivers than to lower-IQ female receivers. However, as discussed in Section \ref{sec:EmpStrategy}, the coefficient estimates related to higher-IQ receivers are biased because low-IQ dictators are more likely to face higher-IQ receivers. Hence, we move to Column 2, which includes dictator IQ fixed effects. Comparing the coefficient estimates of Columns 1 and 2, the only difference is the coefficient estimate on higher-IQ receivers ($\beta_2$), which is in line with the discussion in Section \ref{sec:EmpStrategy}. Indeed, the dictator IQ fixed effects in Column 2 are jointly statistically significant with a p-value of 0.07.

Column 2 shows that the coefficient estimate on female receivers ($\beta_3$) is 0.031, suggesting that male dictators may allocate slightly more to lower-IQ female receivers than to lower-IQ male receivers, albeit the difference is statistically insignificant. Also, the coefficient estimate on higher-IQ receivers ($\beta_2$) is 0.054, suggesting that male dictators may allocate more higher-IQ male receivers than to lower-IQ male receivers, albeit the difference is statistically insignificant. The sum of the coefficient estimates on the interaction term between higher-IQ receivers and female receivers, and on higher-IQ receivers ($\beta_1+\beta_2$) is 0.071, suggesting that male dictators may allocate slightly more to higher-IQ female receivers than to lower-IQ female receivers, albeit the difference is statistically insignificant. The sum of the coefficient estimates on the interaction term between higher-IQ receivers and female receivers, and on female receivers ($\beta_1+\beta_3$) is 0.048, albeit statistically insignificant, suggesting that male dictators may allocate more to higher-IQ female receivers than to higher-IQ male receivers.

However, as discussed in Section \ref{sec:EmpStrategy}, the comparison between allocations to higher-IQ female receivers and allocations to lower-IQ female receivers is confounded because higher-IQ receivers earn more in the IQ test than lower-IQ receivers, which may induce dictators' inequality aversion. Also, the comparison between allocations to higher-IQ female receivers and allocations to higher-IQ male receivers is confounded because men consider women to be more attractive than men and thus may allocate more to female receivers than to male receivers.

To address these concerns, we turn to our coefficient of interest: The coefficient estimate on the interaction between higher-IQ receivers and female receivers ($\beta_1$), which is 0.017 and statistically insignificant, suggesting that male dictators’ allocation differences towards higher-IQ female and male receivers and towards lower-IQ female and male receivers are statistically indistinguishable. Column 3, which controls for dictator and receiver demographics, as well as for round and dictator-receiver social distance to increase statistical power, presents quantitatively similar empirical patterns to Column 2.\footnote{Appendix Table D1 presents results where I gradually add controls and show that the main results are not driven by specific controls.}\textsuperscript{,}\footnote{Appendix Table D2 presents results where the standard errors are neither cluster-robust nor heteroskedasticity-robust to see how the standard errors change by clustering and making it robust to heteroskedasticity.}\textsuperscript{,}\footnote{Appendix Table D3 presents the same results as Table \ref{tab:MainRegMale} but with female dictators. While the coefficient estimates on higher-IQ receivers and on female receivers are slightly negative albeit statistically insignificant, the coefficient estimate of the interaction between higher-IQ receivers and female receivers remains positive, albeit statistically insignificant.} Thus, the results are inconsistent with the hypothesis that men are less generous to higher-IQ women.

A caveat is that the confidence interval is not very tight. Looking at the 95\% confidence intervals of the interaction between higher-IQ receivers and female receivers ($\beta_1$), our coefficient of interest, we can reject at the 5\% significance level that the effect size is no smaller than -8.9 to -10.1 percentage points and no larger than 13.4 to 15.1 percentage points. As references, \textcite{engel_dictator_2011} finds via meta-analysis that male dictators allocate 14.3 percentage points more to female receivers than to male receivers, and \textcite{dufwenberg_generosity_2006} find using a university-student sample that male dictators allocate 7.6 to 8.9 percentage points more to female receivers than to male receivers. To halve the confidence intervals, we need approximately 1560 participants (=390*4).\footnote{The number is based on the OLS standard error formula.}

\subsection{Distribution Results}

\begin{figure}[!htb]
\begin{center}
\caption{Dictator allocations to higher-IQ female receivers -- Distribution, male dictators}\label{fig:AllocationCDFResidualizedMale}
\includegraphics[width=\columnwidth]{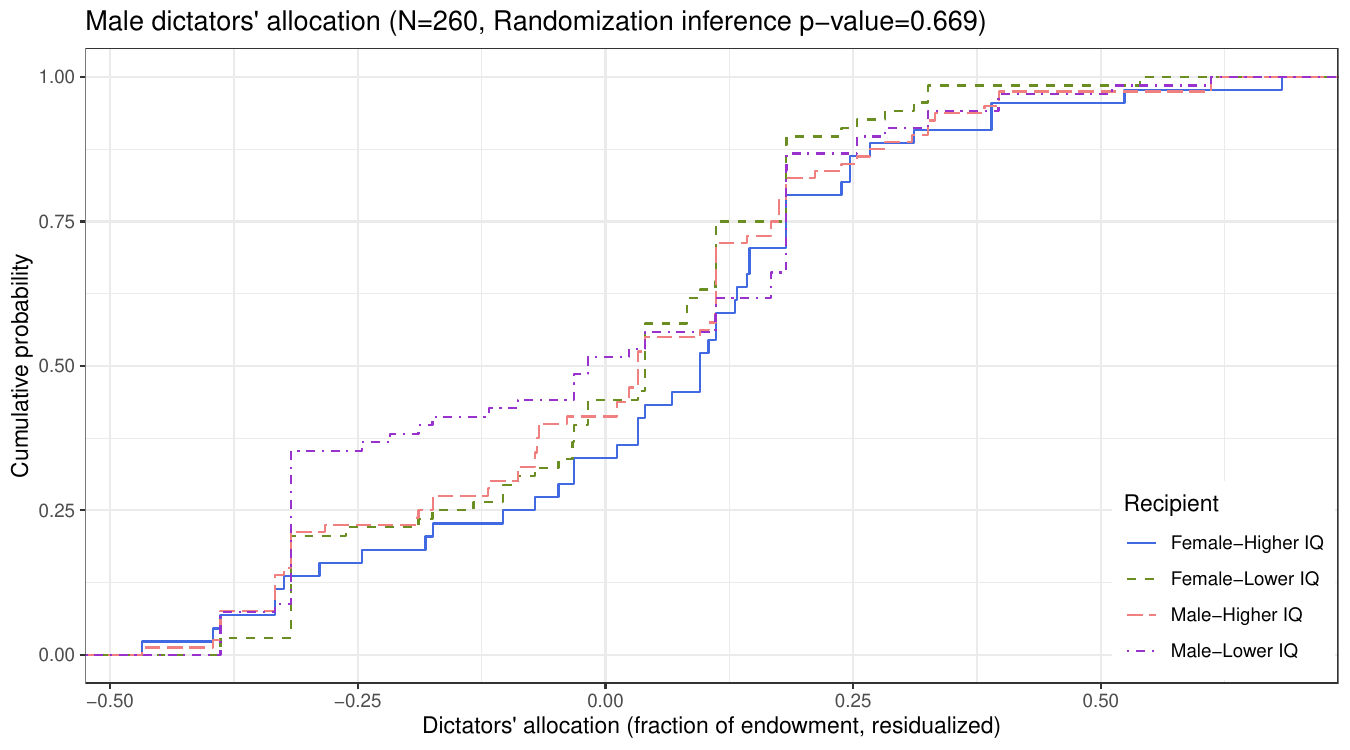}%
\end{center}
\vspace{-12pt}
\begin{tablenotes}[flushleft]
\item \footnotesize \textit{Notes:} The figure presents the empirical CDF of dictator allocations by receiver types, residualized with the dictator-IQ fixed effects, to give a causal interpretation to the differences. The randomization inference p-value \parencite{young_channeling_2019} is calculated with the Kruskal-Wallis test with 2000 random draws. I use randomization inference to address arbitrary dependency among allocations. The null hypothesis is that all CDFs coincide.
\end{tablenotes}
\end{figure}

Although we did not find support for the hypothesis in mean, we may find support in distribution. To see this, Figure \ref{fig:AllocationCDFResidualizedMale} presents the empirical CDF of dictator allocations by receiver types, residualized with the dictator-IQ fixed effects to give a causal interpretation to the differences,\footnote{Residualized allocation is residual from a regression of dictator allocations on dictator-IQ fixed effects.} and shows that the results in Section \ref{sec:RegResults} also hold in distribution. It shows that across almost the whole dictators' allocations, male dictators allocate slightly more to higher-IQ female receivers than to lower-IQ male receivers, higher-IQ male receivers, and lower-IQ female receivers, albeit the differences are statistically insignificant (the randomization inference p-value is 0.669, which is robust to arbitrary dependency among allocations, calculated with the Kruskal-Wallis test with 2,000 random draws).\footnote{Appendix Figure D3 presents the same results as Figure \ref{fig:AllocationCDFResidualizedMale}, but with female dictators and shows essentially the same empirical patterns.}

Thus, the results are inconsistent with the hypothesis that men are less generous to higher-IQ women even in distribution; rather, if anything, men are slightly more generous to higher-IQ women than to higher-IQ men.

\subsection{Robustness Check}
\subsubsection*{``Beauty Premium''}
In the main results, I control for the possibility that dictators allocate more to more attractive receivers -- the so-called ``beauty premium'' \parencite{rosenblat_beauty_2008} -- by taking double differences. Yet, I further address this concern in Column 4 of Table \ref{tab:MainRegMale}. In addition to all the controls in Column 3, Column 4 controls for the receivers' attractiveness via fixed effects (adding dummies for each level of attractiveness ratings, which are jointly significant with a p-value of 0.001), and shows that the magnitude of the coefficient of interest, the interaction term between higher-IQ receivers and female receivers ($\beta_1$), remains essentially the same as in Column 3, albeit the standard errors are larger due to a drop in the sample size by 23\%.\footnote{44 out of 193 receivers, or 23\% of all receivers, refused to show their photos in the follow-up experiment.}

\subsubsection*{Inequality Aversion due to Differential Earnings in the IQ Test}
I control for dictators' inequality aversion \parencite{fehr_theory_1999} due to the differential earnings in the IQ test between higher-IQ receivers and dictators via IQ fixed effects. Yet, it assumes that male dictators believe higher-IQ female and male receivers solved the same number of IQ test questions. Thus, I examine the validity of this assumption in Column 5 of Appendix Table D1. Instead of allocation data for male dictators, Column 5 uses data for male receivers' beliefs on the IQ levels of the other receivers.\footnote{See Section \ref{sec:Experiment} for the justification that it is a valid proxy of the male dictators' beliefs about the receivers' IQ level.} It shows that male receivers believe that higher-IQ female receivers solve about 0.82 more IQ test questions (=1.109-0.291) or earn about 0.41\texteuro\ more (about 6.5\% of the dictator endowment) than higher-IQ male receivers. Thus, if anything, the main results are underestimated in the absence of inequality aversion.

\subsubsection*{IQ Rank Differences}
Dictators see the difference between the receivers' and their own IQ rank. Yet, in the main results, I only consider whether the receivers' IQ rank is higher than the dictators' IQ rank. To investigate the heterogeneity by the difference in the IQ rank, I estimate the following equation:
\begin{equation}\label{eq:RankDiff}
\begin{split}
Allocate_{ij} = &\sum_{k=-5,\ne 1}^5\beta_1^k \mathbbm{1}[ReceiverIQRank_{j}-DictatorIQRank_{i}=k]\times FemaleReceiver_j \\
&+ \sum_{k=-5,\ne 1}^5\beta_2^k \mathbbm{1}[ReceiverIQRank_{j}-DictatorIQRank_{i}=k] + \beta_3 FemaleReceiver_j \\
&+ X_{ij}'\gamma + \mu_i^{IQ} + \epsilon_{ij}
\end{split}
\end{equation}
where $ReceiverIQRank_{j}$ is receiver $j$'s IQ rank, $DictatorIQRank_{i}$ is dictator $i$'s IQ rank, and other variables are as defined in equation \ref{eq:MainEq}. In the main specification, $HigherIQReceiver_{ij}$ is equal to $\sum_{k=-5}^{-1}\mathbbm{1}[ReceiverIQRank_{j}-DictatorIQRank_{i}=k]$ and the omitted category is $\sum_{k=2}^{5}\mathbbm{1}[ReceiverIQRank_{j}-DictatorIQRank_{i}=k]$.

\begin{figure}[!htb]
\begin{center}
\caption{Dictator allocations to higher-IQ female receivers -- IQ rank differences, male dictators}\label{fig:RegRankDifMale}
\includegraphics[width=.8\columnwidth]{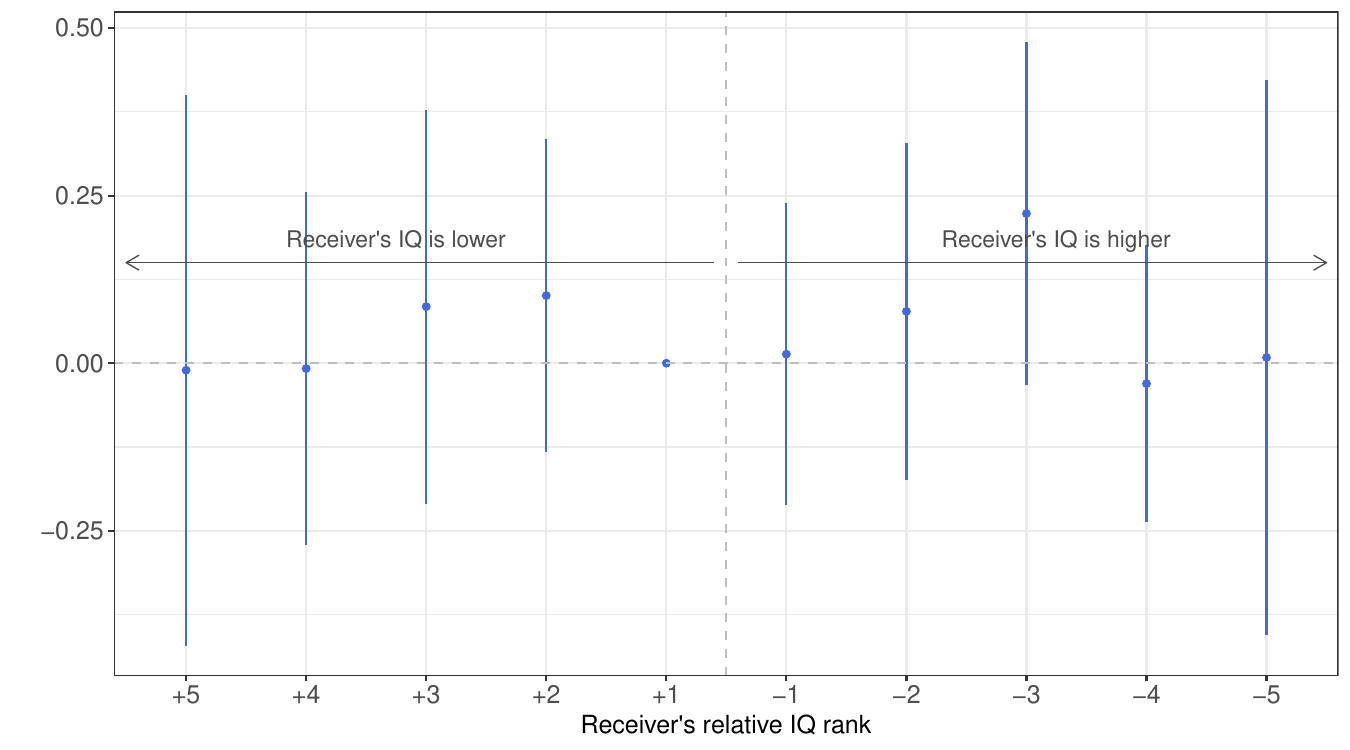}%
\end{center}
\vspace{-12pt}
\begin{tablenotes}[flushleft]
\item \footnotesize \textit{Notes:} This figure plots the OLS estimates of $\beta_1$s in equation \ref{eq:RankDiff}, along with the 95\% confidence intervals. The omitted category is the receiver's relative IQ rank is 1. The standard errors are clustered at the dictator level with \textcite{pustejovsky_small-sample_2018}'s small cluster bias adjustment.
\end{tablenotes}
\end{figure}

Figure \ref{fig:RegRankDifMale} plots the OLS estimates of $\beta_1$s of equation \ref{eq:RankDiff}, along with the 95\% confidence intervals. We do not see any statistically or quantitatively significant heterogeneity due to the IQ rank differences -- all $\hat{\beta}_1$s are close to each other. While the estimate of $\beta_1^{-3}$ is larger than the other estimates, it is not statistically significantly different from the omitted category, and there is no consistent pattern when the receiver's IQ is higher than that of the dictator's as the estimates of $\beta_1^{-4}$ and $\beta_1^{-5}$ are very small.

\subsubsection*{Other Concerns}
\begin{figure}[!htb]
\begin{center}
\caption{Dictator allocations to higher-IQ female receivers -- Sub-sample analysis, male dictators}\label{fig:RobustRegPlotMale}
\includegraphics[width=.8\columnwidth]{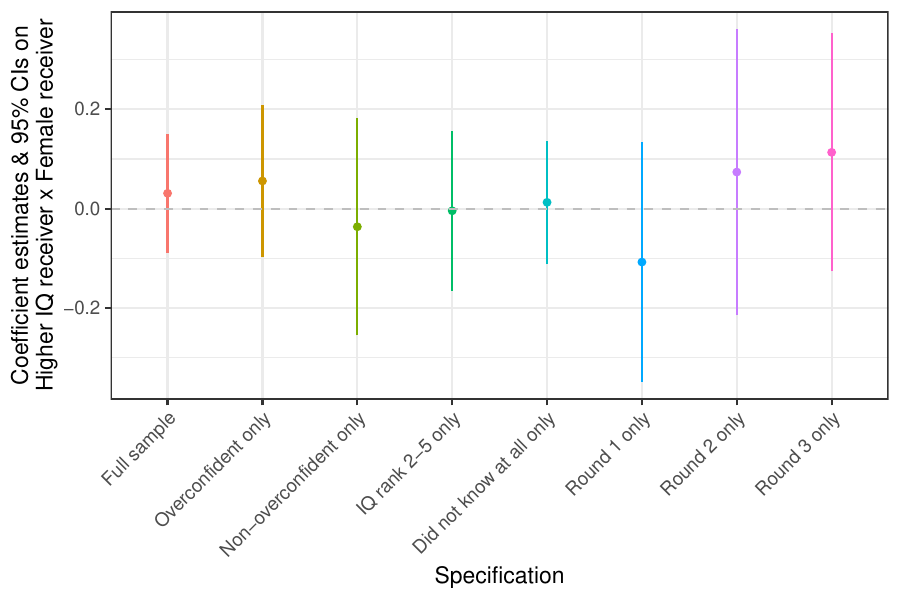}%
\end{center}
\vspace{-12pt}
\begin{tablenotes}[flushleft]
\item \footnotesize \textit{Notes:} This figure presents the OLS estimates of $\beta_1$ and their 95\% confidence intervals of equation \ref{eq:MainEq} with the same controls in Column 3 of Table \ref{tab:MainRegMale} but with sub-samples of male dictators. ``Full sample'' is the same estimate as in Column 3 of Table \ref{tab:MainRegMale}, provided as a reference. The standard errors are clustered at the dictator level with \textcite{pustejovsky_small-sample_2018}'s small cluster bias adjustment for specifications from ``Full sample'' to ``Did not know at all only'' and heteroskedasticity-robust with \textcite{bell_bias_2002}'s small sample bias adjustment for specifications ``Round 1 only,'' ``Round 2 only,'' and ``Round 3 only.''
\end{tablenotes}
\end{figure}

Figure \ref{fig:RobustRegPlotMale} presents the OLS estimates of $\beta_1$ and their 95\% confidence intervals of equation \ref{eq:MainEq} with the same controls in Column 3 of Table \ref{tab:MainRegMale} but with sub-samples of male dictators. ``Full sample'' is the same estimate as in Column 3 of Table \ref{tab:MainRegMale}, provided as a reference. As we see, the point estimates are quantitatively very close to those of the full sample, although the confidence intervals are wider due to a drop in the sample size. One noticeable heterogeneity is round effects, where male dictators allocate less in round 1 and more in rounds 2 and 3. This heterogeneity could possibly be due to the experimenter demand effect, as dictators may have realized that the experiment was about gender and IQ.

\section{Conclusions}\label{sec:Conclusion}
The literature finds that men are more generous to women than to men. Still, there is scarce evidence about how men's higher generosity towards women interacts with skills important for one's career success, such as cognitive skills. The literature suggests that men's higher generosity towards women may be due to paternalism and thus be reversed if women excel in those skills. To fill the gap in the literature, this paper uses a dictator game to examine whether men are less generous to women when those women excel in cognitive skills. Although male dictators allocate more to attractive receivers, consistent with the so-called ``beauty premium,'' and male dictators with high IQs (or those who earn more on the IQ test) allocate less, I exogenously vary the receivers' IQ relative to that of the dictators and use difference-in-differences to control for these associations.

The results, however, are inconsistent with the hypothesis; if anything, male dictators give more to higher-IQ female receivers than to higher-IQ male receivers relative to the difference in giving between lower-IQ female and male receivers. The results hold both in mean and distribution, and female dictators' allocation patterns are qualitatively similar to male dictators. The caveat is that the confidence intervals are not very tightly estimated. Taken together, men are no less generous to women even when women excel in cognitive skills, and thus, women who excel in cognitive skills may not receive less favorable treatment than equally intelligent men in the labor market.

One possible explanation for the results would be that the gender norm violations are offset by something else; for example, it is possible that women are more likely to appreciate other people's work, women with high IQs do so more effectively, and men (and women) reciprocate indirectly. Indeed, \textcite{folke_workplace_2023} show that workplaces with a higher female-worker ratio have a higher incidence of appreciation for others’ work. Another potential explanation is that gendered attitudes develop as people get older, and university students -- my sample -- have not formed gendered attitudes yet. If this is the case, intelligent women may still receive less favorable treatment than equally intelligent men in the labor market. These two explanations are not mutually exclusive, but testing these explanations is beyond the scope of this paper. Nevertheless, my paper showed that, at least among the university student population in Italy, intelligent women would not receive less favorable treatment than equally intelligent men.

\newpage
\clearpage
\begin{singlespace}
\printbibliography
\end{singlespace}

\end{document}